\documentstyle[11pt,newpasp,twoside,epsf]{article}
\def\edcomment#1{\iffalse\marginpar{\raggedright\sl#1\/}\else\relax\fi}
\reversemarginpar

\begin{document}
\title{Blazar jets: the spectra}
\author{Gabriele Ghisellini}
\affil{Osservatorio Astron. di Brera --- V. Bianchi 46, I-23807 Merate, Italy}

\begin{abstract}
The radiation observed by blazars is believed to originate 
from the transformation of bulk kinetic energy of relativistic 
jets into random energy.
A simple way to achieve this is to have an intermittent central power 
source, producing shells of plasma with different bulk Lorentz factors.
These shells will collide at some distance from the center, producing
shocks and then radiation.
This scenario, called {\it internal shock model}, is thought to be at 
the origin of the $\gamma$--rays observed in gamma--ray bursts 
and can work even better in blazars.
It accounts for the observed key characteristics of these objects,
including the fact that radiation must be preferentially produced
at a few hundreds of Schwarzschild radii from the center, but 
continues to be produced all along the jet.
At the kpc scale and beyond, the slowly moving parts of a (straight) 
jet can be illuminated by the beamed radiation of the core, while 
the fast parts of the jet will see enhanced cosmic microwave radiation.
In both cases the Inverse Compton process can be the dominant radiation
process, leading to a copious production of high energy (X--rays and beyond)
radiation in both radio loud quasars and radio--galaxies.
\end{abstract}

\section{Introduction}

We believe that the radiation we see from blazars comes from the
transformation of bulk kinetic into random energy of particles,
which then produce beamed emission.
How to produce the large velocities of the plasma in the jet and 
which is the dissipation mechanism are still a matter of debate, 
but there is no doubt that nature succeds in producing collimated 
outflows with bulk Lorentz factors $\Gamma\sim$ 5--20 for blazars, 
and even higher for gamma--ray bursts.
Only in recent years we began to estimate the power of jets,
through the radiation they emit (e.g. Celotti \& Fabian, 1993)
and especially through the energy required to be transported to the lobes 
(Rawling \& Saunders, 1991).
It has been found that the observed jet radiation must be a small fraction
of the total energetics, even of the last decade witnessed a factor 10
increase in the power observed to be emitted by blazars as a class,
thanks to the high energy $\gamma$--ray observations of EGRET, onboard
the Compton Gamma Ray Observatory satellite.
\begin{figure}
\plotone{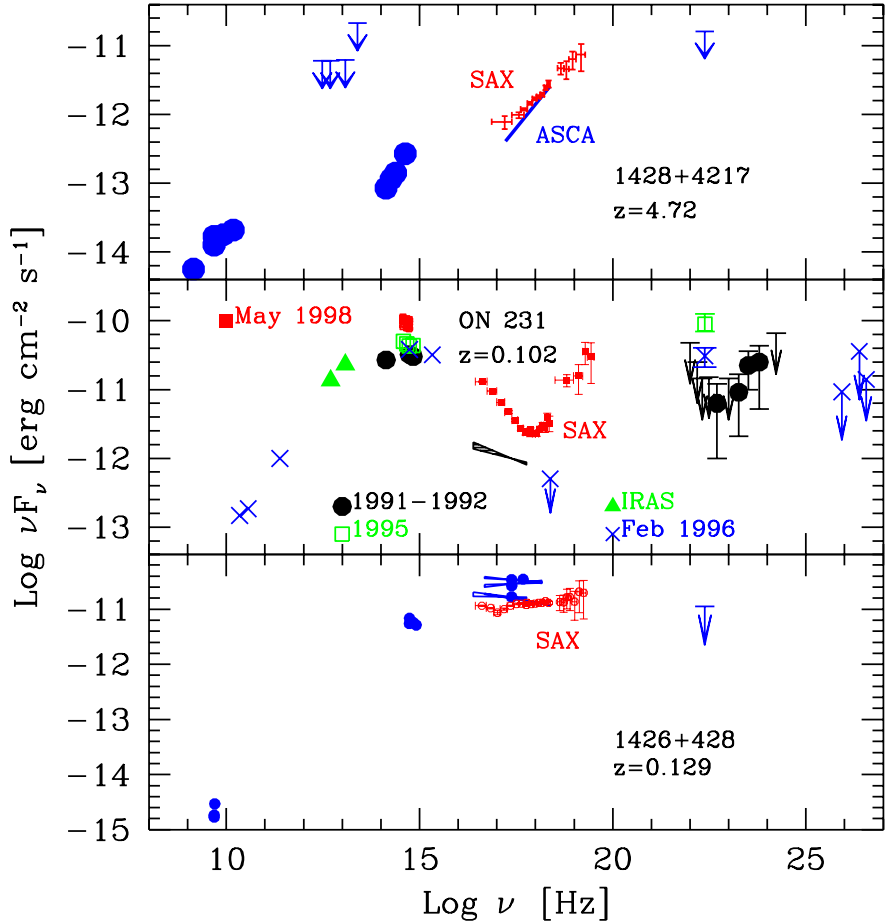}
\caption{Three examples of SED of blazars to illustrate the blazar sequence.
The top panel shows the most distant radio--loud AGN known, at $z=4.72$
(from Fabian et al. 2000). Its luminosity in hard X--rays exceeds,
if isotropic, $10^{49}$ erg s$^{-1}$. 
Note that the peak of the inverse Compton emission is in the MeV band,
and that the hard X--ray emission dominates the power output.
In the mid panel we show the intermediate BL Lac object ON 231, in which
the synchrotron emission dominates the steep soft X--ray
flux, and a very flat inverse Compton component dominates
above a few keV (from Tagliaferri et al., 2000).
In the bottom panel we show the SED of the extreme BL Lac 1426+428,
in which the synchrotron component peaks above 100 keV (from Costamante
et al., 2000). This is the third example of BL Lac object with a 
synchrotron peak located above 100 keV, besides Mkn 501 and 1ES 2344+514.
In these low luminosity class of sources, the emitting
electrons can attain the highest energies, making these objects
the best candidates to be detected in the TeV band.
Note the broad band X--ray range of {\it Beppo}SAX and how
useful it is to characterize the SED in all three classes of blazars.
}
\end{figure}
The EGRET observations, and the detection of a few sources 
(Mkn 421, Mkn 501, 2344+514 and PKS 2155--304) in the
TeV band by ground based Cherenkov telescopes, renewed the interest 
about blazars, allowing the discovery that their Spectral Energy Distribution 
(SED) is characterized by two broad peaks, whose location is a function
of the observed bolometric luminosity (Fossati et al. 1998, 
Ghisellini et al. 1998).
These peaks have been interpreted as due to synchrotron and inverse Compton 
radiation, respectively.
Blazars form a well defined sequence, with low powerful objects having
both peaks at a similar level of luminosity, and located at higher 
frequencies than in more powerful objects, in which the inverse 
Compton peak dominates the emission.
In Fig. 1 we show three examples of SED of blazars with different power, 
to illustrate the overall behavior and what can be the contribution 
of X--ray observations in these three classes of objects.
Recent observations of high redshift ($z>4$) blazars (Celotti, these 
proceedings), and of low power BL Lacs (Costamante et al. these proceedings) 
have extended the blazar sequence at both ends, confirming the original trend.

At the high luminosity end of the sequence we find interesting 
lower limits on the bulk kinetic power that the jet can carry,
requiring  it to be larger than the power dissipated in radiation,
derived dividing the apparent luminosity (assuming isotropy and no beaming) 
by the square of the bulk Lorentz factor.
Results indicate that jet of FSRQ (flat spectrum radio quasars)
must have a large kinetic power (Celotti, these proceedings).
As an example, PKS 0836+710 has an apparent luminosity of $10^{49}$ erg 
s$^{-1}$, which requires a jet power of at least 
$10^{47}/\Gamma_1^2$ erg s$^{-1}$ (Tavecchio et al., 2000a).
Note that in these sources most of the jet power must not be dissipated
through radiation, but must feed the extended radio structures.

At the low luminosity end of the blazar sequence we find objects whose
synchrotron spectrum peaks in the X--ray band, indicating very large
energies of the emitting electrons.
Here we can learn about the acceleration mechanism, and find
good candidates to be detected in the TeV band
(Costamante et al, these proceedings).

Here I will focus on two main topics, namely how the 
{\it internal shock scenario} can explain the main characteristics
of blazars, and how the large (and very large) scale
X--ray jets recently observed by $Chandra$ can be interpreted.

\section{Internal shocks}

The key idea of the internal shock scenario is to assume a
central engine working intermittently, i.e. producing discrete
blobs or shells of plasma moving at slightly different velocities.
In this case there will always be a later faster shell catching 
up a slower earlier one.
If the initial separation of the two shells is $R_0$ and
the Lorent factors $\Gamma$ differ by a factor 2, the 
collision will take place at $R\sim R_0\Gamma^2$.

This idea is not new: Rees (1978) proposed it to explain some 
features of the M87 jet, by it was almost forgotten in the AGNs 
field, even if it became the leading scenario to explain the
$\gamma$--ray radiation of gamma--ray bursts.

\subsection{Points in favor}

\subsubsection{``Low" efficiency}

Consider two shells with bulk Lorentz factors $\Gamma_1$ and $\Gamma_2$
and mass $m_1$ and $m_2$.
Conservation of energy and momentum implies that a fraction
$\eta$ of the total bulk kinetic energy must be dissipated:
\begin{equation}
\eta \,=\, 1-\Gamma_f\, { m_1+m_2\over \Gamma_1 m_1 +\Gamma_2m_2}
\end{equation}
where $\Gamma_f=(1-\beta_f^2)^{-1/2}$ is the bulk Lorentz factor 
after the interaction and is given through 
(see e.g. Lazzati, Ghisellini \& Celotti 1999)
\begin{equation}
\beta_f = {\beta_1\Gamma_1m_1+ \beta_2\Gamma_2m_2 \over
\Gamma_1m_1+ \Gamma_2m_2}
\end{equation}
The above relations imply, for shells of equal masses and 
$\Gamma_2=2\Gamma_1=20$, $\Gamma_f=14.15$ and $\eta=5.7\%$.
The fraction $\eta$ is not entirely available to produce radiation,
since part of it is in the form of hot protons and magnetic field.
This is the efficiency for a single collision.
Merged shells (that have already collided) can however collide again
with other shells (or merged shells), increasing the total
fraction of kinetic energy transformed into radiation to 5--10\%.
The rest is transported to the outer radio structures of the jet.
The small efficiency in producing radiation is a major problem
in the field of gamma--ray bursts, but is indeed a positive
feature for blazars, since we need to transport most of the power
to the outer radio lobes.

\subsubsection{Right location}

One of the most important implications of the EGRET observations of 
blazars is the realization that most of the luminosity of these 
sources must be emitted in a well localized region of the jet
(Ghisellini \& Madau 1996).
This region cannot be too close to the jet apex, to avoid absorption
of $\gamma$--rays by X--ray radiation produced by the jet itself
or by the accretion disk and its corona.
On the other hand the rapid $\gamma$--ray variability suggests
that the $\gamma$--ray emitting zone is not too far from the 
jet apex.
Hundreds of Schwarzschild radii are indicated.
In powerful blazars,
this distance is conveniently close to the distance of the 
Broad Line Region (BLR), which can produce seed photons
for the formation of the $\gamma$--ray flux.

On the other hand, the entire jet must emit some radiation,
particularly at radio frequencies, where synchrotron self--absorption
limits the emission in the inner part of the jet.
In the internal shock scenario the emission at large scales is due to 
collisions between shells that have already collided once (or more times). 
The efficiency in this case is lower, since the bulk Lorentz factors
have already averaged out somewhat.

\subsubsection{Variability}

Internal shocks are a very simple way to produce variability.
In this scenario there is a typical variability timescale
(at least for the first collisions) connected to the initial 
separation between two colliding shells and their width.
If the initial separation of two consecutive shells is $R_0$, 
they will collide at the distance $R=R_0\Gamma^2$, but the corresponding time 
will be observed Doppler contracted by the factor 
$(1-\beta\cos\theta)\sim 1/\Gamma^2$ and will be of the same order 
of the initial separation $R_0/c$.
The duration of each flare is linked to the duration of the collision, 
which will be of the order of the shock crossing time.
Inhomogeneities within the shells and small scale instabilities, 
if present, can produce variability on shorter timescales.

\subsubsection{Correlated variability}

High frequency emission is mainly produced by shells colliding
for the first time at $R\sim 10^{17}$ cm from the jet apex.
Lower frequency (radio and far IR) flux is produced further out
in the jet, when merged shells collide with other merged shells.
Therefore there should be some correlations between the light curves
at different frequencies, especially between the $\gamma$--ray and
optical fluxes and the mm--radio flux.

\subsection{Internal shocks: a powerful blazar}

We (Spada et al. 2000) simulated the case of a powerful
blazar jet, of average bulk kinetic power of $10^{48}$ erg s$^{-1}$,
carried by shells or blobs injected in the jet, on average, every
few hours, with a bulk Lorentz factor chosen at random in the
range [10--30].
The shell width is initially of the same order of the initial shell--shell
separation. 
Material in the shell (both protons and electrons) are assumed to be initially
cold, and consequently the shell is assumed to have a constant width
until the first collision takes place.
After that, the shell width is assumed to expand with the sound velocity.
The first collisions happen at a few$\times 10^{16}$ cm, well within
the Broad Line Region (BLR), assumed to be located at $5\times 10^{17}$ cm
and to reprocess 10\% of the disk luminosity, assumed to be equal to  
$10^{46}$ erg s$^{-1}$.
\begin{figure}
\plotone{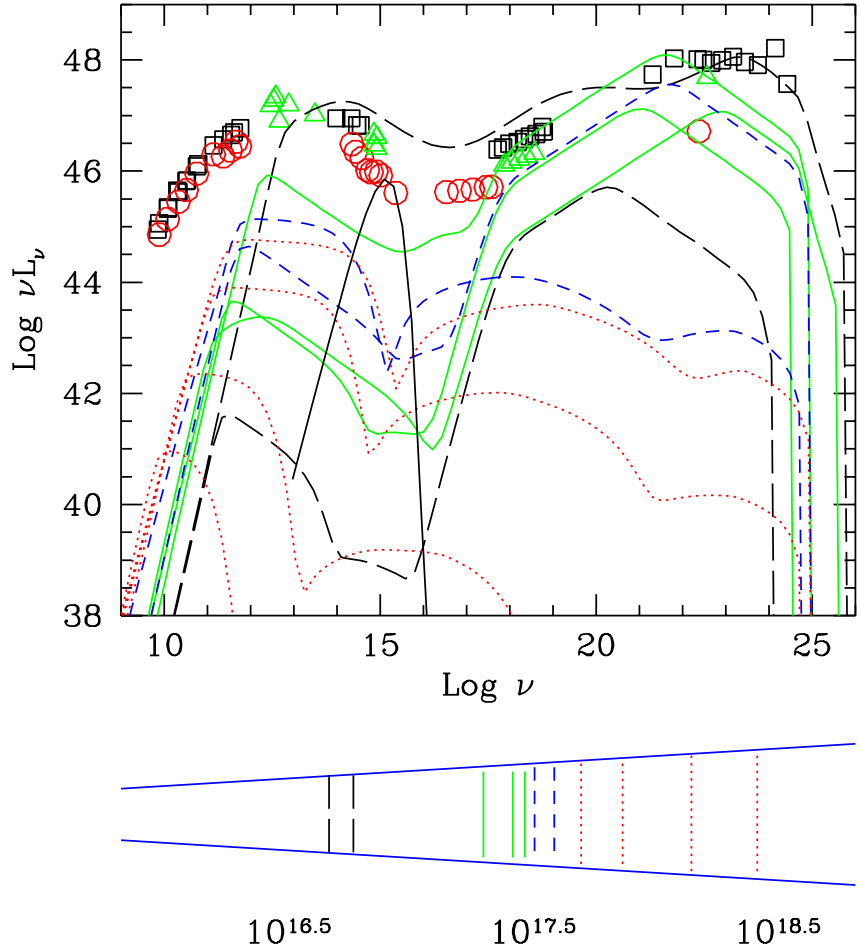}
\caption{Some spectra calculated in the internal shock scenario,
produced at different jet locations, as labelled in the lower panel.
For illustration, we have superimposed the SED of 
3C 279 during three simultaneous observing campaigns
(i.e. in 1991, 1993 and 1996, see Maraschi et al. 1994 and
Wehrle et al., 1998).
The blackbody peaking at $10^{15}$ Hz is the assumed spectrum
of the accretion disk. 10\% of this luminosity is reprocessed
by the BLR located at $5 \times 10^{17}$ cm.}
\end{figure}
For simplicity, a fixed and constant fraction of the energy dissipated 
during each collision is assumed to go into the electron and to the magnetic 
field components.
The emitting relativistic particles are assumed to have a broken power--law 
energy distribution throughout the entire emitting zone.
This energy distribution is derived by assuming to inject in the source 
electrons with a single power law distribution
with minimum electron energy $\gamma_{\rm min}$ whose 
value is found by energetic considerations.
Limits in computing time do not allow us to consider details of
spectral changes on a timescale faster than the light crossing
time of a single shell (few hours when $R\approx 10^{17}$ cm
and a month on the parsec scale).

Particles emit by synchrotron,
synchrotron self--Compton (SSC), 
and Compton scattering off the 
external radiation (EC) produced by the BLR. 

We simulate the evolution of the total spectrum summing the locally 
produced spectra of those regions of the jet which are simultaneously active 
in the frame of the observer. 
In Fig. 2 we show some spectra, each corresponding to one particular 
shell--shell collisions at a different distance, as 
sketched in the bottom panel. 
The entire time dependent evolution 
can be seen in the form of a movie at the URL:
{\tt http://www.merate.mi.astro.it/$\sim$lazzati/3C279/index.html}.

As can be seen, the predicted spectra are extremely variable
(more so at the higher frequencies).
First collisions are the most efficient in converting bulk energy
into radiation, since in this case the ``$\Gamma$--contrast": 
(i.e. $\Gamma_2/\Gamma_1$) is the largest.
These collisions, taking place inside the BLR, make
the inverse Compton process the most important cooling agent.
The corresponding spectrum therefore peaks in the $\gamma$--ray band.
Collisions taking place outside the BLR (preferentially
between shells that have already collided once) have
a smaller $\Gamma$--contrast, and see relatively less seed photons.
This makes the synchrotron component to dominate
(dashed and dotted lines in Fig. 2).

In Fig. 2 we also show the observational data of 3C 279 during three
observational campaigns.
While the agreement is gratifying, we stress that we
have not yet tried to obtain ``a best fit" for this source.
We have rather used fiducial numbers for the initial time separation 
of the shells, their initial width and the overall average bulk kinetic
energy of the jet.

We have performed a cross--correlation analysis of the light curves
at different frequencies.
As expected, there is no lag between the $\gamma$--ray 
and the X--ray and the optical fluxes, which are mainly
produced in the same (inner) zones of the jet.
Instead, there is a well defined delay  of $\sim$40 days 
between the $\gamma$--ray and the far infrared (1 mm) fluxes.
This is easily explained by the fact that the mm radiation is
preferentially produced at some pc from the center, yielding
a time delay of
\begin{equation}
\Delta t = {{\Delta R} \over {c \Gamma^2}} \sim 38.5 \;
\Delta R_{19} \; \Gamma_1^{-2} \;\; {\rm days}, 
\end{equation}
\begin{figure}
\plottwo{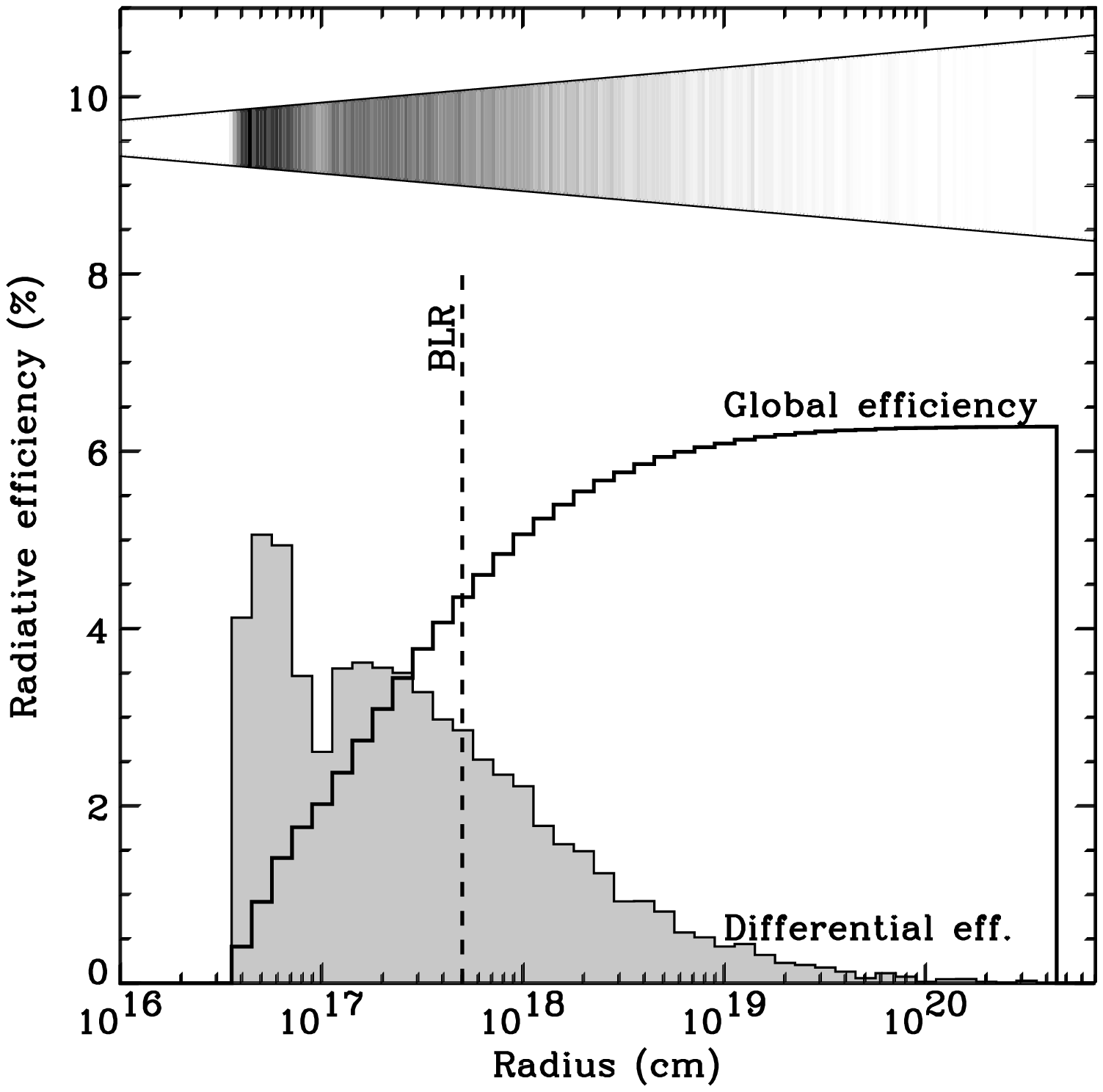}{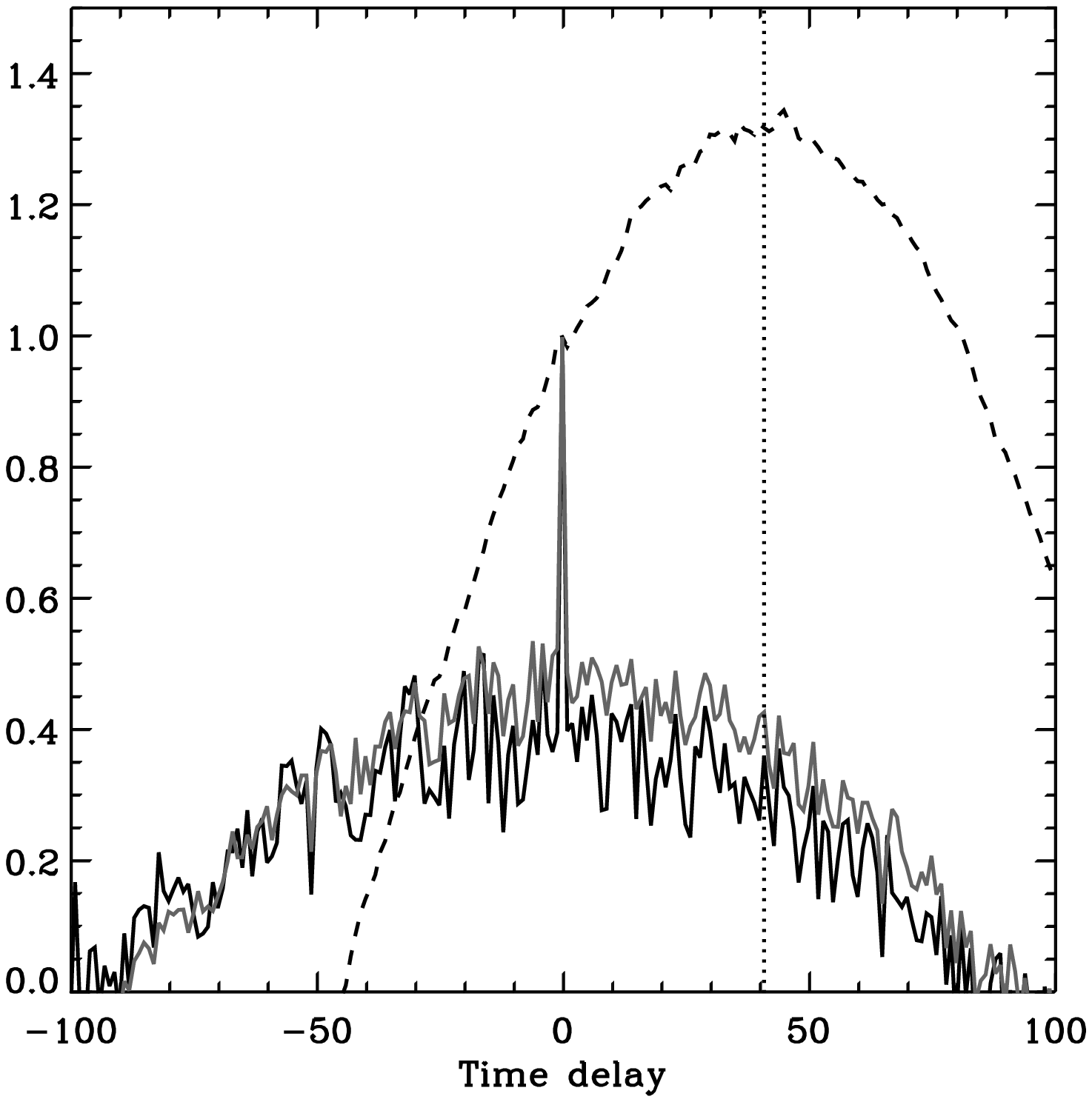}
\caption{{\it Left:}
The radiative efficiency versus the collision radius
for the particular simulation of Fig. 2.
The solid line refers to the global efficiency, i.e. the fraction 
of the total kinetic energy of the wind radiated on scales smaller 
than a given radius; the shaded histogram shows instead the differential 
efficiency, i.e. the fraction of bulk kinetic energy radiated at a 
given radius interval (multiplied by a factor of 10 for clarity). 
The cone at the top shows a grey--tone representation of the 
differential efficiency of the jet. 
The darker the color the higher the efficiency.
{\it Right:} Cross correlation of the simulated light curves,
between $\gamma$-- and X--rays, $\gamma$--rays and optical, and
$\gamma$--rays and the mm band (short dashed line).
 }
\end{figure}
One of the main assumptions of our simulations has been to
calculate the value of the magnetic field considering only
the energy dissipated in each collisions, and neglecting
any seed magnetic field which, surviving from
previous collisions, can be amplified by shock 
compression. 
As a consequence, the magnetic field $B$ scales with distance
$R$ (from the jet apex) as $B\propto R^{-1.5}$, resulting in very
small magnetic field values on the outer zones.
In turn this implies long cooling times and small variability
in the radio band.

While we hope to ``cure" this in future work, we would like to stress
that there are other possibilities for enhanced dissipation at 
large distances.
The relativistic plasma could in fact be shocked by ``obstacles"
in the jet, or it can interact with the (steady) walls of the jet.
This case resembles what in the gamma--ray burst field is called
``external shock scenario", in which the collisions are much 
more efficient in converting bulk into random energy than internal 
shocks (i.e. the shell decelerates much more).
In this case it is natural to expect some velocity structure
across the jet, with a fast ``spine" along the axis and slower
``layers" along the borders.
Indeed, Chiaberge et al. (2000) found evidences
for this structure, for explaining the spectra of radiogalaxies.

\section{Chandra jets}

We have shown (Celotti, Ghisellini \& Chiaberge 2000) that if some
dissipation takes place in the jet at distances greater than 1--100 kpc,
then the Inverse Compton scattering process with external radiation 
is favored with respect to the synchrotron self Compton
process, leading to an 
X--ray flux larger than what expected by a pure SSC model.
This study was motivated by the recent detection by Chandra of large
scale X--ray jets both in radio--galaxies and in quasars, especially
in PKS 0637--752 (Chartas et al. 2000; Schwartz et al. 2000).
This source is particularly interesting because radio VSOP 
observations detected superluminal motion in the (pc scale) 
jet (Lovell 2000), which implies $\Gamma >17.5$ and a viewing angle 
$\theta< 6.4$ degrees. 
This in turn implies a de--projected X--ray jet length of almost a Mpc.
We have proposed that, at these scales, the jet is still relativistic
$\Gamma=10$--15) and then its randomly accelerated particles can 
efficiently interact with the cosmic microwave background (CMB), 
producing beamed X--ray rays through the inverse Compton process.
\begin{figure}
\plottwo{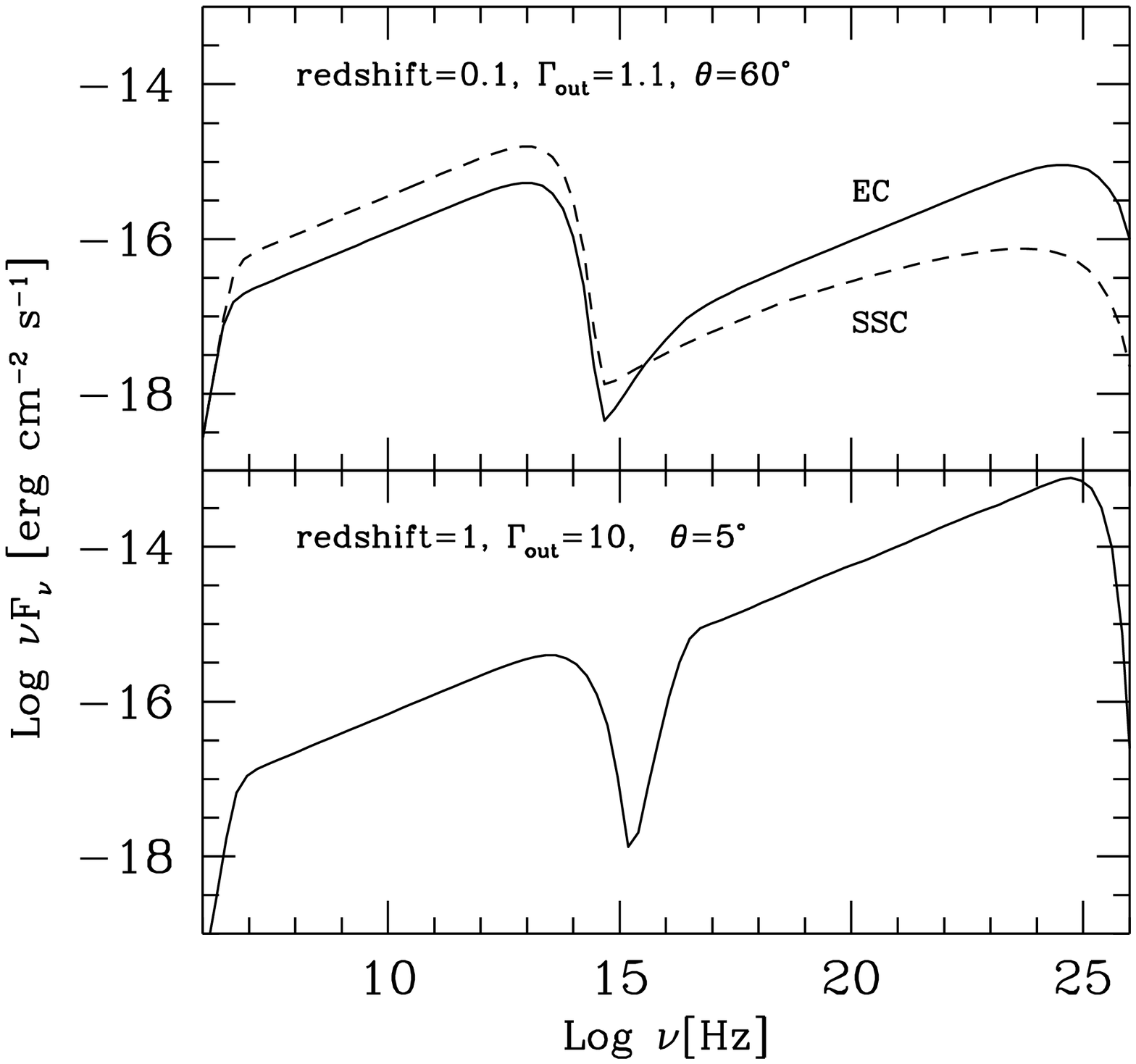}{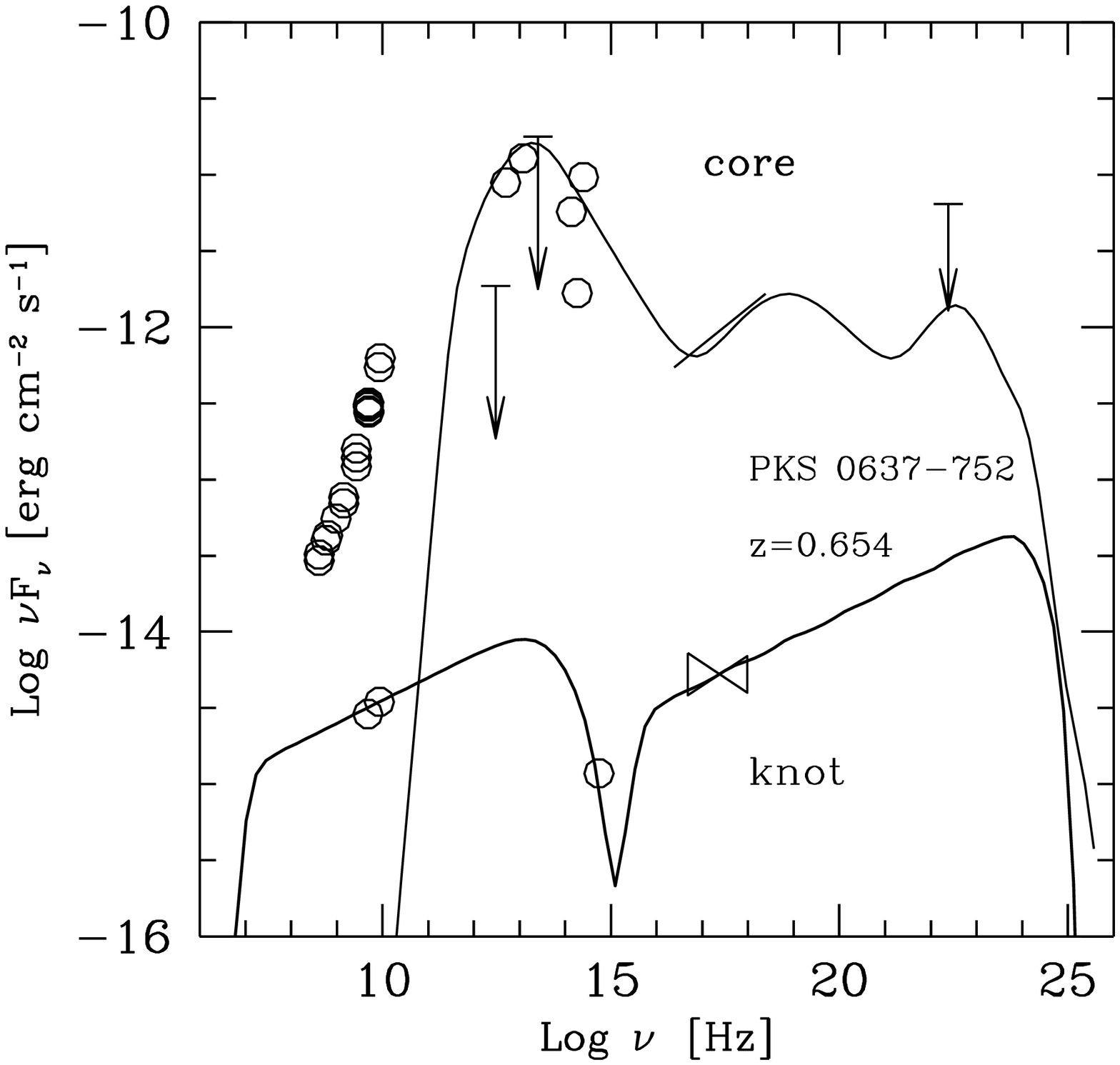}
\caption{
{\it Left:} SED calculated assuming that at distances of 10 kpc from 
the center a region of 1 kpc of size embedded in magnetic field of 
$10^{-5}$ G radiates an intrinsic power of $3\times 10^{41}$ erg s$^{-1}$.
The nuclear (blazar) component emits an intrinsic power of $10^{43}$ 
erg s$^{-1}$. 
The {\it upper panel} shows the emission from a layer with 
$\Gamma_{\rm layer}=1.1$ and viewing angle $\theta=60^\circ$.
The dashed line corresponds to the SED assuming that electrons
emit SSC radiation only.
The solid line takes into account the radiation field coming from the 
core of the jet and illuminating the region.
The {\it bottom panel} shows the emission from a spine moving with
$\Gamma_{\rm spine}$ at a viewing angle $\theta=5^\circ$.
{\it Right:} The SED of the core and the large scale knot of PKS 0637--752,
together with the models for both components (solid lines). From 
Celotti, Ghisellini \& Chiaberge, (2000).}
\end{figure}

It may seem unusual to invoke relativistic bulk motion at these
extremely large scales. But:

i) If the jet decelerates, it has to dissipate, and a sizable
fraction of the initial energy must be radiated, contrary
to the requirement that most of the jet power survives
up to the radio structures. 
More so if the jet becomes only mildy relativistic, through, e.g. 
entrainment. 

ii) Suppose that the 100 kpc jet is only mildly relativistic.
The produced radiation is then only marginally beamed,
{\it enhancing} the requirements on the total energetics
with respect to a relativistic jet (Ghisellini \& Celotti, 
2000, in prep.).

Celotti, Ghisellini \& Chiaberge (2000) were able to fit both
the core and the large scale jet emission of PKS 0637--752
(see Fig. 4), with a bulk Lorentz factor of 17 for the core and 14 
for the jet, {\it conserving the bulk kinetic power of the jet} between the
two emission sites, and with nearly equipartition between
magnetic and electron energy densities in the large scale 
jet emission site (see Tavecchio et al. 2000b for another solution).

In the frame comoving with the jet, the energy density of the cosmic 
background radiation is $\propto \Gamma^2(1+z)^4$. 
Therefore distant blazars with fast large scale jets should be even brighter 
than PKS 0637--752 in X--rays with respect to their synchrotron 
radio--optical components, making the efforts of Chandra to detect them easier.

As mentioned, besides fast ``spines" we can have
slowly moving ``layers", and also these components can 
emit more inverse Compton radiation with respect to a pure
SSC model, because the layers can be illuminated by
the beamed radiation produced by the core,
if the core and the large scale jets are aligned.
Fig. 4 (top panel) shows one example, with the comparison
with a pure SSC spectrum.
Being slow, and possibly only mildly relativistic, the layers produce 
radiation which is much more isotropic than the spine: at small
viewing angles (blazar case) this component is outshined by
the spine component, but it can become visible as the viewing
angle increases (i.e. in radio--galaxies).

\section{Discussion}

Internal shocks can dissipate the right amount of jet power at the
right location, originating in a natural way the violent
variability observed in blazars.
This scenario has the virtue to be at the same time simple and 
quantitative, offering a coherent view of almost all 
the radiative jet, from milliparsecs to kiloparsecs.
In the $\gamma$--ray band we expect the most pronounced variability,
in agreement with observations, and we are now working to find how the
$\gamma$--ray duty cycle 
(i.e. the fraction of the time spent in high $\gamma$--ray states)
scales with the initial separation
of the shells and their initial bulk Lorentz factor
to compare it with data coming from 
the foreseen $\gamma$--ray satellites AGILE and GLAST.

We hope that this scenario will also explain the observed
``blazar sequence", linking the overall blazar spectrum with
the jet power (i.e. the bulk motion power).
We also hope, with longer simulations, to be able to see if there
is a relation between the strongest $\gamma$--ray flares and the 
birth of radio superluminal blobs at the pc scale.

The emission predicted by the internal shock scenario (as it is now) beyond 
the kpc scale likely underestimates the X--ray flux (both soft and hard). 
At these distances the energy densities in magnetic field and locally
produced synchrotron radiation are very small, and seed photons
of the cosmic microwave background on one hand and of the core of the
jet on the other hand
are important contributors to the inverse scattering process
for fast spines and slow layers, respectively.

\end{document}